# Forward modeling of coronal mass ejection flux ropes in the inner heliosphere with 3DCORE

**C. Möstl[1]\*, T. Amerstorfer[1], E. Palmerio[2], A. Isavnin[2], C. J. Farrugia[3], C. Lowder[4,5], R.M. Winslow[3], J. M. Donnerer[1], E. K. J. Kilpua[2], P.D. Boakes[1]**

[1]Space Research Institute, Austrian Academy of Sciences, Graz, Austria.
[2]Department of Physics, University of Helsinki, Helsinki, Finland.
[3]Institute for the Study of Earth, Oceans, and Space, University of New Hampshire, Durham, NH, USA.
[4]Department of Mathematical Sciences, Durham University, Durham, United Kingdom.
[5]Southwest Research Institute, Boulder, CO, United States.

\*Corresponding author: Christian Möstl (christian.moestl@oeaw.ac.at)

**Key Points:**

- A new semi-empirical model (3DCORE) for simulating the propagation of coronal mass ejection magnetic flux ropes is introduced.

- 3DCORE is able to model the observations of a coronal mass ejection on 2013 July 9–13 observed at MESSENGER and Wind.

- Solar, coronagraph and MESSENGER observations are used as constraints, resulting in a good match of the modeled and observed *Dst* index.

**Abstract**

Forecasting the geomagnetic effects of solar storms, known as coronal mass ejections (CMEs), is currently severely limited by our inability to predict the magnetic field configuration in the CME magnetic core and by observational effects of a single spacecraft trajectory through its 3D structure. CME magnetic flux ropes can lead to continuous forcing of the energy input to the Earth's magnetosphere by strong and steady southward-pointing magnetic fields. Here, we demonstrate in a proof-of-concept way a new approach to predict the southward field *Bz* in a CME flux rope. It combines a novel semi-empirical model of CME flux rope magnetic fields (*3-Dimensional Coronal ROpe Ejection* or *3DCORE*) with solar observations and in situ magnetic field data from along the Sun-Earth line. These are provided here by the MESSENGER spacecraft for a CME event on 2013 July 9–13. 3DCORE is the first such model that contains the interplanetary propagation and evolution of a 3D flux rope magnetic field, the observation by a synthetic spacecraft and the prediction of an index of geomagnetic activity. A counterclockwise rotation of the left-handed erupting CME flux rope in the corona of 30 degree and a deflection angle of 20 degree is evident from comparison of solar and coronal observations. The calculated *Dst* matches reasonably the observed *Dst* minimum and its time evolution, but the results are highly sensitive to the CME axis orientation. We discuss assumptions and limitations of the method prototype and its potential for real time space weather forecasting and heliospheric data interpretation.

# 1 Introduction

We have recently seen the emergence of novel techniques to describe the evolution of coronal mass ejections (CME) from the Sun to the Earth by combining CME parameters derived from observations with geometrical and physics-based approaches, hence they are appropriately called "semi-empirical" models. They either model the full propagation of the CME magnetic flux rope (MFR) and its deformation in the solar wind [*Isavnin*, 2016] or use solar observations to set the type of MFR [e.g. *Bothmer and Schwenn*, 1998; *Mulligan et al.* 1998; *Marubashi et al.* 2015, *Palmerio et al.* 2017] and subsequently simulate the Earth's trajectory through the structure [*Savani et al.* 2015, 2017; *Kay et al.* 2017]. They can be seen as tools similar to the forward modeling of the CME in coronagraphs [*Thernisien et al.* 2009], but aimed instead at producing synthetic in situ observations. One crucially important aspect is that such approaches allow the long-lead time prediction of the southward magnetic field component of the interplanetary magnetic field *Bz*, preferably in *Geocentric Solar Magnetospheric* (GSM) coordinates. Southward *Bz* is the prime requirement for geomagnetic storms as it opens the subsolar magnetopause by magnetic reconnection allowing efficient transfer of energy, plasma and momentum to the magnetosphere [e.g., *Dungey* 1961]. During quiet solar wind intervals, *Jackson* et al. [2015] showed the possibility to predict *Bz* variations of a few nT by a combination of a near-Sun magnetic field modeling technique with interplanetary scintillation. The *Bz* component is steady and strong in CME MFRs and thus they drive the strongest geomagnetic storms [*Huttunen et al.* 2005, *Zhang et al. 2007*]. Accurate CME *Bz* predictions are currently not possible, but the ordered magnetic fields in MFRs can be seen as a key that nature has given us to be able to forecast very strong geomagnetic storms that could be very harmful to humankind [e.g. *Oughton et al. 2017*]. We just need to figure out how to use this key.

Semi-empirical techniques simulate the CME evolution in a much simpler and computer efficient way than full 3D MHD solar wind models like Enlil, which include the interaction of the CME with the background wind [e.g. *Odstrcil et al.* 2004]. There have been various efforts to include the MFR in numerical simulations of CME eruption and evolution [e.g. *Manchester et al.* 2008], but those cannot yet be carried out in real time and do not yet give very accurate Bz forecasts [review by *Manchester et al.,* 2017]. Several efforts are currently underway to include the CME flux rope structures in numerical simulations for real time predictions, but results are not yet available. Semi-empirical models have two main strengths: (1) the models can be run on any computer and are immediately suitable for predicting a space weather event, even in real time, and (2) the researcher is able to quickly adapt the model output to observations, narrowing the range of free parameters and enhancing future predictions based on the model. The researcher has thus a strong control of the model output, at the expense of having many free parameters that need to be set in order to produce realistic results [e.g., *Isavnin,* 2016].

This paper introduces a new semi-empirical method that we call *3-Dimensional COronal Rope Ejection* or *3DCORE*. This is the first such model that contains the interplanetary propagation including deceleration, expansion, the measurement by a synthetic spacecraft at any given heliospheric location and production of a geomagnetic index from the simulation. It is designed in such a way that it is aimed at real-time problem solving and does not contain a description of the 3D MFR that is completely physically correct. In contrast to the aforementioned models, we place 2.5D cross sections in the desired shape and do not employ a complete 3D MFR solution [e.g. *Hidalgo et al.* 2002, *Nieves-Chinchilla et al.* 2016]. This is essentially a work-around for the problems that are associated with deforming a 3D physical solution of CME MFRs. Here, by putting magnetic field cross sections in the desired shape, we follow an inverse way of looking at this problem compared to *Isavnin* [2016] and *Wood et al.* [2017], who fill an initial envelope shape with magnetic field lines. It is currently unclear which method produces the most consistent results with observations. Thus, there is a need to develop several versions of semi-empirical 3D MFRs to be able to figure out the advantages and disadvantages of each model and their ability to act as a tool to interpret heliospheric observations as well as to predict CME effects in real time.

We can now test all these methods extremely well with observations. The European Union HELCATS project (www.helcats-fp7.eu) has collected solar wind data sets and brought them into easy-to-use formats from 2007 to present. The HELCATS products include not only in situ data at multiple points by the missions Venus Express, MESSENGER, STEREO and Wind, but also solar imaging, coronagraph and heliospheric imaging along with modeling like Thernisien et al. [2009] and Davies et al. [2012]. Details on the catalogs that contain these observations can be found in Möstl et al. [2017]. Also concerning the upcoming missions *Solar Orbiter*, *Parker Solar Probe, BepiColombo* and the *Cubesat for Solar Particles* (*CuSP*), semi-empirical models will likely form very valuable tools in heliophysics research for many years to come.

## 2 Method

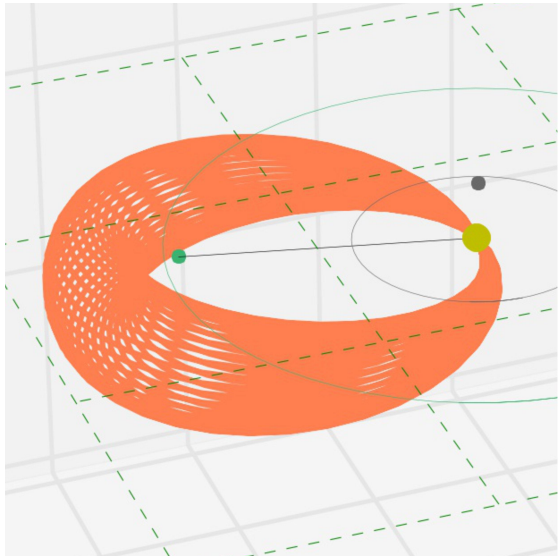

**Figure 1: 3DCORE prototype geometry**. *The model envelope (orange) consists of a tapered torus that is attached to the Sun at all times. The global shape as well as the cross-section are circular. The Sun is shown as a yellow circle (not to scale) and the Earth is shown as a green dot.*

**Figure 1** shows the 3DCORE geometry (for the exact mathematical formulation, please see the supplementary material). The model in its first prototype form has these basic assumptions: (1) it consists of 2.5D Gold-Hoyle [e.g. *Farrugia et al.* 1999, *Hu et al.* 2014] circular cross sections forming the uniformly twisted flux rope in 3D as a global torus, with a global circular shape attached to the Sun. Note that the global circular geometry resembles the so called "Harmonic Mean" approach in the field of heliospheric imaging [*Lugaz,* 2010] if the flux rope axis is not inclined to the solar equatorial plane. The axial field has a value $B_0$, and the twist number $\tau$ is in the range of 1–20 turns per AU [*Hu et al.* 2014]. A 2.5D geometry means that the axial component can vary within the cross-section, but the magnetic field is invariant along a direction orthogonal to the plane of the cross-section. Everything is calculated in *Heliocentric Earth Equatorial Coordinates* (HEEQ). (2) The circular torus is tapered, so the cross-section has a varying radius along the torus, with its largest cross-section extension at the torus apex that then decreases on both sides towards the Sun towards zero. (3) The grid is variable. In this study it contains 10° steps in the $\psi$ coordinate along the torus, so 36 cross sections in total, 10° steps along $\Phi$ around the axis, and between 1 and 20 points along the cross-section radius, with constant distances of 0.01 AU in between radial points along the cross-section radius. The total grid size thus varies for each simulation step: near the Sun there are about 1.3k grid points, which rises to 25k grid points beyond 1 AU. (4) The MFR is rotated to the given latitude, longitude and orientation calculated by rotations with the Euler-Rodrigues formulas. (5) The kinematics of the nose are calculated with the drag-based model [*Vršnak et al.* 2013]. (6) The rest of the MFR moves according to self-similar expansion, with a constant angular width. Thus, the speed of the nose is scaled to each point of the CME. (7) The MFR axial magnetic field declines with distance with a power law of −1.64 [*Leitner et al.* 2007], and does not vary along the $\psi$ direction. (8) The torus cross-section diameter increases following an almost linear expansion law [*Leitner et al.* 2007]. The source code in Python is available for download in Section 6.

In practical applications, the launch time, initial CME speed and the constant direction (latitude and longitude in HEEQ) are derived either from STEREO/HI or COR observations. The MFR handedness and axial field direction is derived from the CME source region magnetograms and extreme ultraviolet images. A subsequent global rotation of the flux rope in the corona is possible to any desired final orientation, which yields all the MFR types known from in situ observations [*Bothmer and Schwenn, 1998*; *Mulligan et al., 1998*]. During the CME outward propagation in the simulation, a virtual spacecraft observes the CME MFR by detecting the nearest magnetic field value in the simulation, which has to be under a certain distance threshold (here set to 0.05 AU). If no point of the MFR grid is close to the virtual in situ spacecraft below the threshold, the flux rope is not detected. These synthetic magnetic field components are then converted from *HEEQ* to *Geocentric Solar Magnetospheric* (GSM) coordinates. This is done based on the methods described in *Hapgood* [1992]. For Earth, the synthetic in situ magnetic field and speed output is then fed into the models by *Burton et al.* [1975] and *O'Brien and McPherron* [2000]. Thereby, a time series of the global magnetospheric *Disturbance storm time* (*Dst*) index is produced. Another, more sophisticated technique for this task was demonstrated by *Temerin and Li* [2006], which due to its complexity will be coupled to 3DCORE in future updates.

## 3 Data

On 2013 July 9 14 UT, a large filament eruption occurred in the northern hemisphere on the Earth-facing side of the solar disk, in a region of an otherwise quiet Sun. The dispersed magnetic fields accompanying the filament are a remnant of an active region that first appeared in April 2013. The subsequent slow CME impacted both MESSENGER, which was 7° away from the Sun-Earth line, and the Sun-Earth L1 point, with a CME transit time of 73 hours, measured from its first appearance in STEREO/SECCHI/COR2 on 2013 July 9 16 UT to the shock arrival at L1 on 2013 July 12 16:47. This is a single CME event that lead to a moderate geomagnetic storm at Earth. At L1, around the time of CME impact, the solar wind speed was stable around 400 km s$^{-1}$, and a high-speed stream impacted the Earth only about 6 days after the CME, so this event is free of interactions from other CMEs or high-speed streams.

**Figure 2** shows the solar observations. In Fig. **2a**, in images in the *Solar Dynamics Observatory* (SDO) *Atmospheric Imaging Assembly* (AIA) 171 Å channel [*Lemen et al.* 2012], coronal loops are seen in the northern hemisphere, accompanied by a large filament in SDO/AIA 304 Å (**Fig. 2b**), spanning from northeast to the center of the solar disk. The filament spine clearly follows an inverse S-shape, which is the signature of a magnetic flux rope with left-handed chirality [see summary in *Palmerio et al.* 2017]. Thus, the filament follows the hemispheric chirality rule. **Fig. 2c** shows the photospheric magnetogram by the SDO *Helioseismic and Magnetic Imager* (HMI) [*Scherrer et al.* 2012]. The leading westward polarity of photospheric magnetic fields underlying the filament channel  is negative, consistent with Hale's law for solar cycle 24 [e.g. *Pevtsov et al.* 2014]. The axial field of such a left-handed flux rope, where the poloidal field is formed by arcades connecting the polarities on either side of the PIL, is expected to point to the southwest, as indicated by an arrow. The filament eruption starts on 2013 July 9 14 UT, and flare ribbons along the entire filament length are visible (**Fig. 2b**). The polarity inversion line (PIL) and the post-eruption arcades (not shown) are straight and are both inclined at a position angle of 220° ± 5°. The position angle is measured counterclockwise from 0° pointing north, 90° to solar east in the solar equatorial plane, 180° to south and 270° to solar west.  The type of magnetic flux rope that is later observed in situ is thus expected to be either west-south-east (WSE) or north-west-south (NWS) [after *Bothmer and Schwenn,* 1998; *Mulligan et al.* 1998].

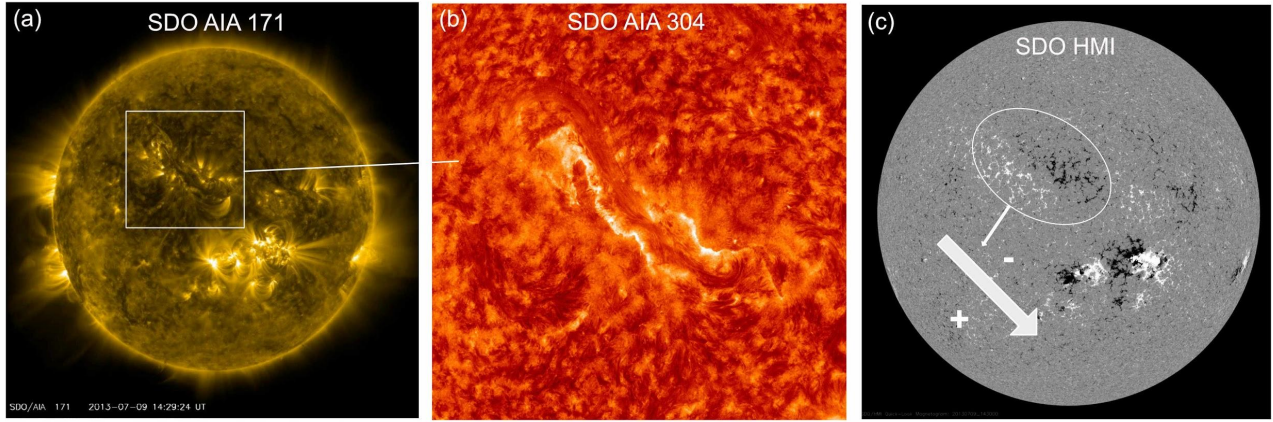

***Figure 2: Overview of solar observations.*** *(a) SDO/AIA 171 Å on 2013 July 9 14:29:24 UT, with the inverse-S shape of the filament indicated, pointing to a left-handed field chirality. (b) Zoom-in on the filament channel in the northern hemisphere in SDO/AIA 304 Å on 2013 July 9 15:29:08 UT, with visible flare ribbons brightening along both sides of the polarity inversion line. (c) SDO/HMI magnetogram on 2013 July 9 14:30 UT, with the MFR axial field of the expected flux rope shown as an arrow.*

In **Figure 3**, about 3 hours after the filament eruption, a CME is visible in both STEREO/SECCHI/COR2 coronagraphs [*Howard et al.* 2008], pointing to solar west in STEREO-Behind imagery and to solar east in STEREO-Ahead, indicating an Earth-directed eruption. Graduate cylindrical shell modeling [GCS, *Thernisien et al.* 2009], **focusing on the CME void and** including both STEREO and the SOHO/LASCO coronagraph images [*Brueckner et al.* 1995], gives an approximately linear speed profile with an average speed from 5.6 to 19 solar radii of 575 ± 60 km s$^{-1}$. The longitude (HEEQ) is −1° ± 5° and its latitude −1° ± 5°, so it is directed practically head-on to Earth. Compared with the approximate center of the filament at N20E15, the direction W1S1 means that the CME has been roughly 20° deflected away from the source region, which is a usual magnitude [*Kay et al.* 2016]. All errors quoted are typical for the method. The CME is rather cone-shaped than having a clear rope shape, and its tilt value is −18 ± 6° to the solar equator (equal to a position angle of 252° ± 6°), but due to the near cone-shape this measure is rather not well defined. It thus seems that the erupting flux rope axis directed at a PA of ~220° has rotated by about 30° from the low corona to coronagraph distances of a few solar radii (axis at PA of ~250°).

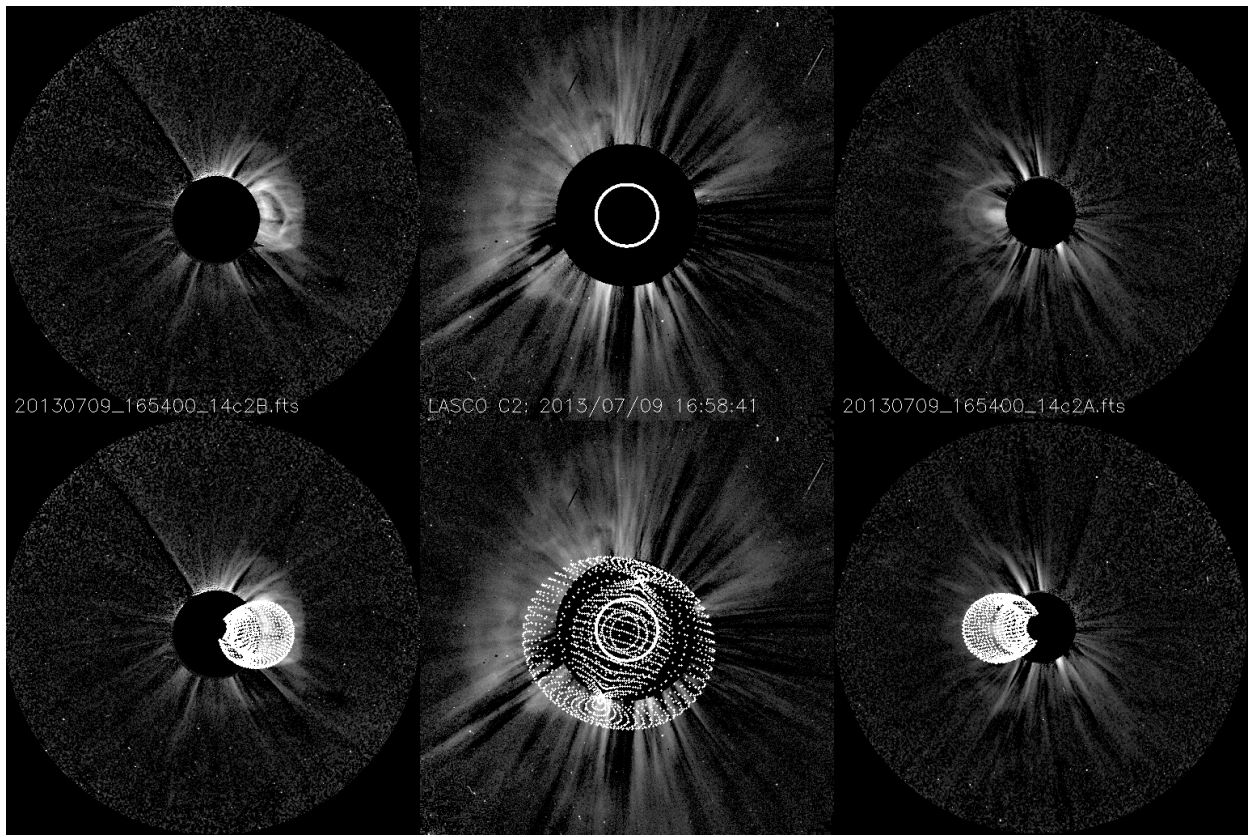

**Figure 3: Coronagraph modeling.** *Upper panels: STEREO-Behind/COR2 (left), SOHO/LASCO/C2 (middle) and STEREO-Ahead/COR2 (right) showing the appearance of the CME in coronagraphs. Lower panels: Graduate cylindrical shell (GCS) model shape overlaid on the same images.*

The heliospheric imager (HI) on STEREO-B [*Eyles et al.* 2009] also observed the event up to about 28.5° away from the Sun, giving a CME direction of −2° ± 10° longitude to the Sun-Earth line by the SSEF30 method, which describes the CME front as a self-similarly expanding circle with 30° half width in longitude [e.g. *Davies et al.* 2012, *Möstl et al.* 2014]. This is a clear Earth-directed CME. The CME propagates symmetrically to the solar equator in north-south direction and does not show a strong deflection in latitude, thus we expect the CME to impact any planets near or in the ecliptic plane, which is close to the solar equatorial plane, along the Sun-Earth line. The CME interplanetary speed from SSEF30 is 513 ± 21 km s$^{-1}$, which forms an average speed for a heliocentric distance of 0.11 to 2.03 AU during the time of the HI observation. A detailed summary of the event is given on the HELCATS webpage referenced in Section 6. Comparing

GCS and HI results shows that GCS is perfectly consistent with the SSEF30 results. Such a convergence between HI geometrical modeling and GCS is typical [*Möstl et al.* 2014], so knowing the result from either method is a good proxy for the results given by the other technique.

**Figure 4** shows the situation in the heliosphere in early July 2013. The arcs in the left plot show modeled CME fronts based on heliospheric imager observations, and the plots on right give an overview of available in situ interplanetary magnetic field data. The CME modeling is again provided by the SSEF30 technique. The CME event we study is the blue circle near Earth (which itself is the green dot), at the movie frame time 2013 July 13 00:00 UT which is shortly before Earth impact. In the arrival catalog ARRCAT, also provided by HELCATS and referenced in Section 6 and described in *Möstl et al.* [2017], the CME shock is predicted to impact MESSENGER at Mercury at a heliocentric distance of 0.4548 AU on 2013 July 11 04:28 UT, and to arrive at the L1 point near Earth on 2013 July 13 01:30 UT at 1.0066 AU. Indeed, in the right part of Figure 4 total magnetic field enhancements are seen in the panels of the MESSENGER and Wind data.

In the ICMECAT, the catalog of interplanetary in situ CME observations [*Möstl et al.* 2017], there is an ICME at MESSENGER starting on 2013 July 11 01:05, which is only about 3 hours prior to the SSEF30 predicted arrival time. This event was originally cataloged in situ by *Winslow et al.* [2015]. In the right panel of Figure 4 the event is visible, in spite of data gaps that arise because the periods when MESSENGER was traversing through Mercury's magnetosphere have been removed. The Wind spacecraft at L1 saw an ICME, also included in ICMECAT, with a shock arrival on 2013 July 12 16:47, which is 9 hours earlier than the predicted arrival time. At Wind, the event lasts for over 2 days until the end of 2013 July 14. The differences between predicted and observed arrival times are well below the current lower limits for CME arrival time prediction of about 12 to 17 hours [e.g. *Vršnak et al.* 2014, *Mays et al.* 2015, *Tucker-Hood et al.* 2017, *Möstl et al.* 2017].

Given that there are no other candidate CMEs that explain the event at L1 and the predicted in situ arrival times are well within the error bars, there is an unambiguous connection between the CME observations in the corona, interplanetary space and in situ data. This is of tremendous importance to accurately test CME magnetic flux rope models, because otherwise solar and in situ observations may not be causally related and we may arrive at incorrect conclusions. Heliospheric imaging observations proved very critical in this task as without them linking

CMEs and interplanetary in-situ CME observations can sometimes be very difficult [e.g., *Kilpua et al.*, 2014].

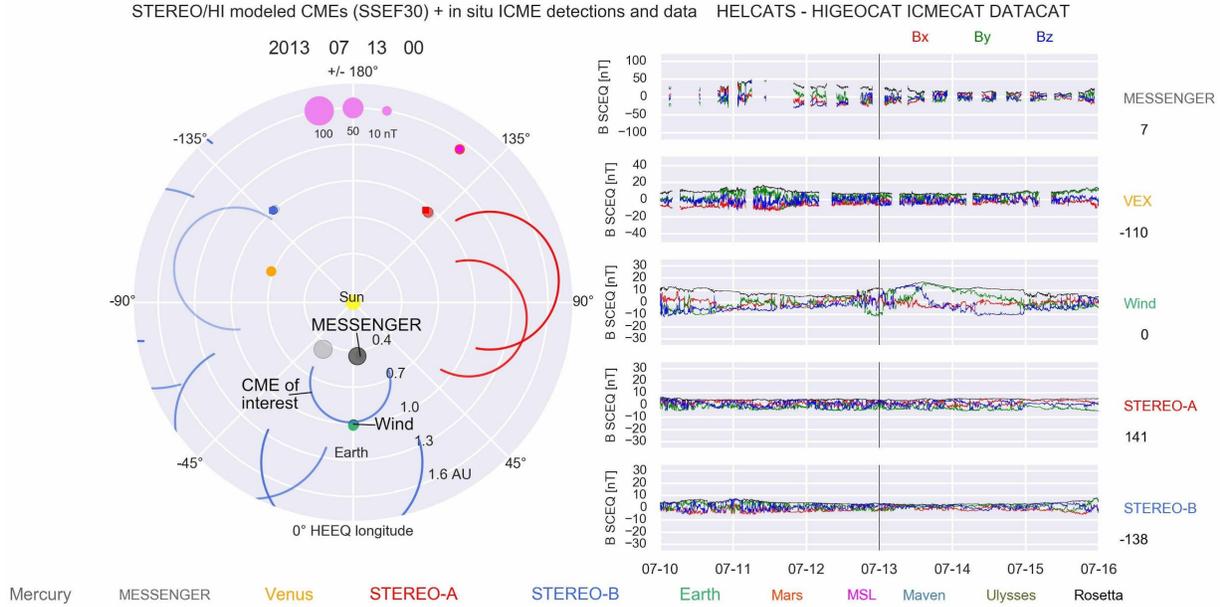

**Figure 4: Overview of interplanetary observations**. *The left panel shows the solar equatorial plane, with planet and spacecraft positions given by the color code at the bottom. Blue and red circles are CME fronts modeled with SSEF30 [Möstl et al. 2014], with the CME in question as a blue circle just before impacting Earth. The other CMEs are to be ignored. On the right panel, in situ magnetic field observations by MESSENGER and Wind show higher total fields and rotations in the components during the CME impact, and no other in situ heliospheric observatory observed the event. Data gaps at MESSENGER are caused by the spacecraft entering the Mercury magnetosphere. This figure is also available as an animation covering 2007 to 2014, see Möstl et al. [2017] and the link in Section 6.*

## 4 Results

We now apply the 3DCORE prototype to the CME event in question. A time resolution of 2 hours is used, and a single run takes about 45 seconds on a personal computer. We use the initial conditions based on GCS, with $575 \pm 60$ km s$^{-1}$ radial speed, the direction longitude and latitude both as $-1° \pm 5°$, and the starting position of the MFR nose on 2013 July 9 20:24 UT at 19.0 solar radii. The position angle of the MFR axis (i.e., the angle measured counterclockwise from the solar north) is 252° ± 6°, as derived from GCS. In this study, we thus assume that a rotation of 32° of the MFR axis from the polarity inversion line inclination to the GCS model orientation took place; MFR axis rotations in this range have been shown to be often consistent with in situ observations [*Marubashi et al.* 2015]. The magnetic field in 3DCORE consists of circular Gold-Hoyle flux rope cross sections which have a uniform twist, set in this study at a value of 5 turns/AU [e.g *Hu et al.* 2014]. The twist value may also be taken from solar observations [e.g. *Vemareddy et al.* 2016] but this is not further pursued here. The MFR has a left-handed chirality, as derived from solar observations in the previous section. The MFR moves according to the drag-based model [DBM, *Vršnak et al.* 2013], and we set the background solar wind speed to 400 km s$^{-1}$, which is the solar wind speed at L1 around the CME launch time. This is of course a rough approximation as it does not include the varying solar wind conditions along the Sun-Earth line. For adding a variability to the background solar wind, the speed values outside of the synthetic MFR are randomly taken from a normal distribution centered around 400 km s$^{-1}$ with a standard deviation of 10 km s$^{-1}$, and the magnetic field at MESSENGER is set to 25 nT (at 1 AU to 5 nT), with a random variation of about 1 nT in the total field.

Now, we have three parameters left that are a priori not well known, the drag parameter $\Gamma$ [in units of $10^{-7}$ km$^{-1}$] which is a part of DBM, the axial magnetic field $B_0$ (in nT) and the MFR diameter $D$ (in AU). In this first prototype study, we constrain these values with MESSENGER data taken near the Sun-Earth line at a heliocentric distance of 0.4548 AU. This is the reason why we have chosen the event as being a radial lineup of two in situ observatories. We are fully aware that such constraints are currently not practicable for real time CME forecasting. However, we have shown the potential for using magnetometer observations near the Sun-Earth line previously from Venus orbit [*Kubicka et al.* 2016], and we need to study how such observations, if available in real time, would enhance space weather forecasts. The idea of using near real time data along the Sun-Earth line for space weather forecasting has a long history [e.g. *Lindsay et al.* 1999], but could become revived soon with the advent of interplanetary small satellites. In the next update of 3DCORE, we will pursue better ways to a priori calculate these 3 parameters by looking at large interplanetary CME statistics for $\Gamma$ and $B_0$ as well as a more sophisticated inclusion of interplanetary CME expansion for calculating $D$ [e.g. *Démoulin et al.* 2009].

The first major unknown is the drag parameter $\Gamma$ in the DBM, which can range from 0.05 to 2 [*Vršnak et al.* 2013, *Temmer and Nitta* 2015]. CME kinematics and subsequent planetary arrival times depend drastically on the $\Gamma$ value. The observed flux rope arrival time at MESSENGER is 2013 July 11 01:57 (55 minutes later than the observed shock arrival). To match this arrival time, $\Gamma$ is set to 1.5 by manually optimization, changing only $\Gamma$ and keeping all other parameters constant. In a next version of the model, a combination with the ElEvoHI model [*Rollett et al.* 2016] could help to find the most appropriate values for $\Gamma$ and the background solar wind speed. The other unknowns concern the diameter $D$ and the axial field strength, $B_0$ of the flux rope, and they are also set by manual optimization to their values **extrapolated (with the power laws)** at 1 AU as $D = 0.24$ AU and $B_0 = 12$ nT;  by this choice the synthetic magnetic field profile at MESSENGER is in approximate accordance with the MESSENGER magnetic field data.

**Figure 5** shows the 3DCORE run with the parameters set as described in comparison to MESSENGER magnetic field data [*Anderson et al.* 2007], with the Mercury magnetic field removed [*Winslow et al.,* 2013; *Winslow et al.* 2015]. Panel (a) is the visualization of the 3DCORE torus. In panel (b), some notable similarities and differences between the observed and simulated magnetic field components are visible. First, the removal of the Mercury magnetosphere is clearly a disadvantage of the chosen CME event, as only a few solar wind data intervals of the flux rope are available. Nevertheless, on the other hand this demonstrates that even very sparse data from along the Sun-Earth line can be used to constrain CME simulations.

The magnetic field component $B_x$ is close to zero, in both the simulation and observation; $B_y$ is positive throughout the simulation but not in the beginning of the observation, whereas at the end of the MFR, on late July 11, the simulated $B_y$ matches the observation. For the intervals when MESSENGER solar wind magnetic field data is available during the CME impact, the simulated $B_z$ component roughly follows the observations. The takeaway message here is a rough consistency between the simulation and observation for the arrival time, total field, and duration when the simulation has been constrained with the MESSENGER observations. Although magnetic field observations from MESSENGER were limited, we can estimate that the flux rope type and orientation that we derived from solar and coronagraph observations are roughly consistent to those observed at Mercury.

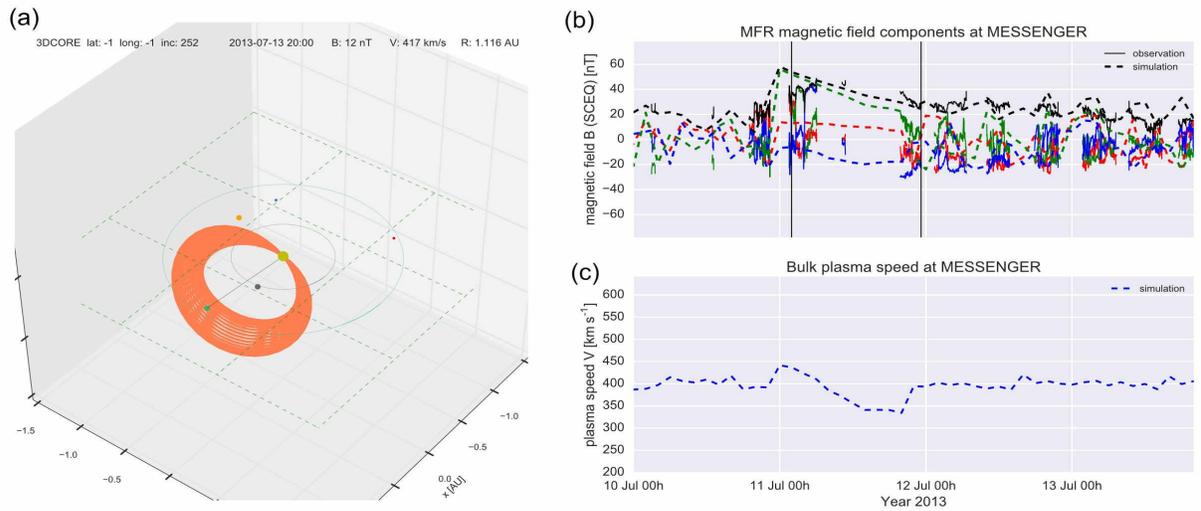

*Figure 5. 3DCORE synthetic magnetic field output for MESSENGER compared to observations*. (a) *Visualization of the 3DCORE torus and spacecraft positions, with MESSENGER as a gray point, Earth as a green dot and the torus extending from the Sun (yellow dot) into interplanetary space.* (b) *The synthetic magnetic field components in SCEQ coordinates (like HEEQ, but longitude corrected for spacecraft): $B_x$ red, $B_y$ green, $B_z$ blue, total field black. Straight lines are observations, dashed lines are the simulation. The magnetic field of Mercury has been removed from the data.* **Two vertical solid lines show the start and end times of the magnetic flux rope in the observations.** (c) *Simulated bulk plasma speed. An animation of panel (a) is available on figshare and youtube, see Section 6.*

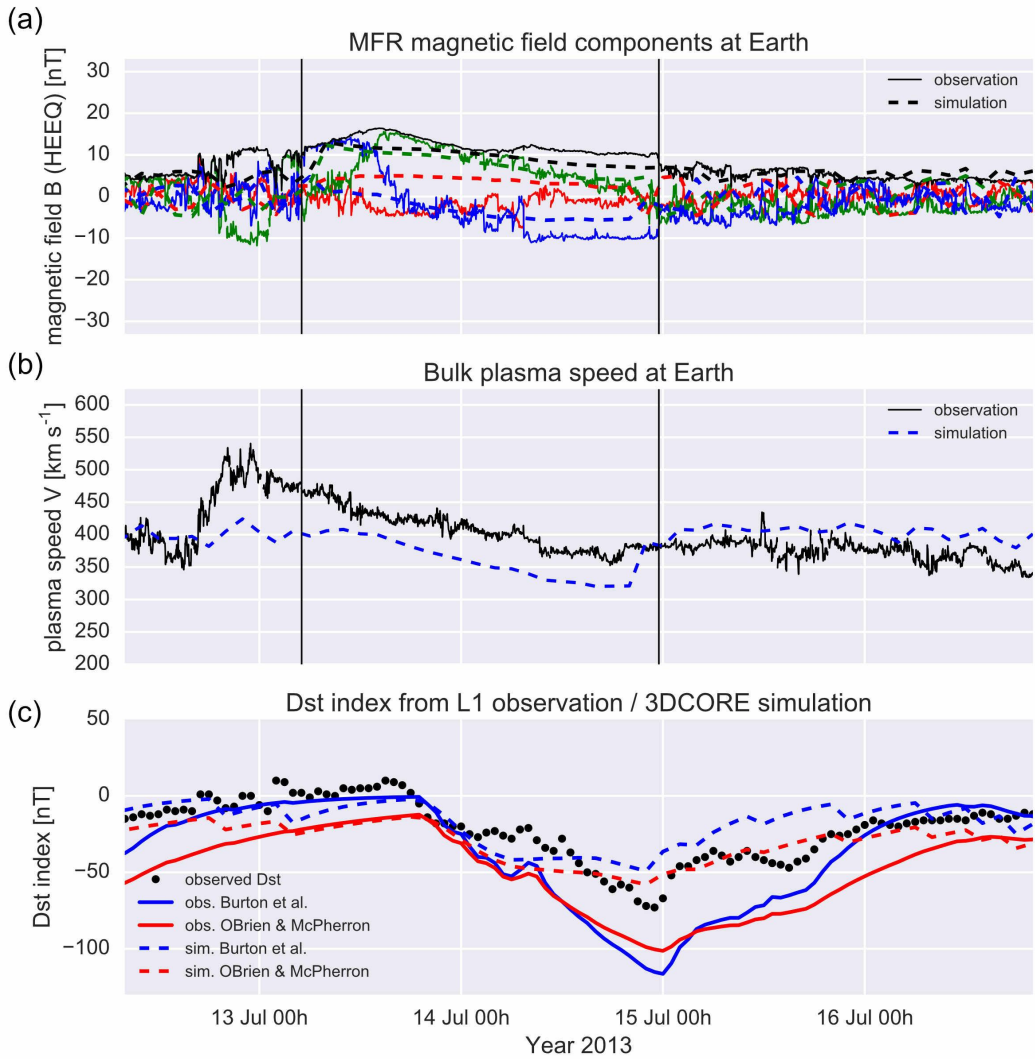

**Figure 6. 3DCORE output compared to Wind observations at Earth / L1 and Dst.** *(a) Synthetic magnetic field components in HEEQ coordinates (dashed lines) compared to observations (solid lines).* **Field components are $B_x$ (red), $B_y$ (green), $B_z$ (blue), and the total field is black. Vertical solid lines delimit the magnetic flux rope in the observations.** *(b) The bulk plasma speed at Wind (solid line) compared to the simulation (dashed line).* **(c) Observed Dst (black dots) and calculated Dst (solid lines) with two methods from OMNI2 speed and magnetic field data (see text). The simulated Dst is represented by the dashed lines, calculated from the 3DCORE speed and magnetic field data at Wind, after conversion of the simulated magnetic field to GSM coordinates.**

**Figure 6** demonstrates the synthetic field with 3DCORE and the observations at Earth L1 by Wind [*Lepping* et al. 1995; *Ogilvie et al.* 1995], and includes the *Dst* index derived from the synthetic field and speed data in comparison to the observed *Dst* taken from the OMNI2 dataset [*King & Papitashvili* 2005]. The CME arrived at Wind on July 12, with an IP shock detected at 16:47 UT. The shock is quite weak, but the rotations typical of a magnetic flux rope structure are very well defined, seen as in **Fig. 6a**. From visual inspection, the rope is of NWS-type (**+$B_z$ at the leading edge, +$B_y$ at the axis, and −$B_z$ at the trailing edge in HEEQ**), thus consistent with the GCS modeling orientation. The consistency between the simulated (dashed lines) and observed (solid lines) is quite **good**. We emphasize that we have set the initial conditions of the simulation with solar observations and coronagraph modeling, and we have constrained 3 currently free simulation parameters with MESSENGER in situ data. There was no information from the Wind spacecraft involved in the modeling, and no fitting process to Wind in situ data was made. Some notable differences between modeling and simulation are: the magnetic field components $B_z$ and $B_y$ is not well modeled at the beginning of the MFR, $B_z$ is underestimated by the simulation near the end of the rope and $B_x$ is slightly positive in the simulation, which is opposite to the observation but not of significant importance for geo-effectiveness. The speed profile in **Fig. 6b**, which is derived from the assumption of self-similar expansion (see supplementary material), is well consistent with the linear decrease of the MFR speed in the observations, but the simulation underestimates the real speed by about 50 km s$^{-1}$.

In **Figure 6c**, we show a comparison of the *Dst* index. The red and blue dashed lines are calculated only with synthetic outputs of the magnetic field components, which are first converted to *Geocentric Solar Magnetospheric (GSM)* coordinates, and with the calculated bulk plasma speed. For conversion of the solar wind to *Dst* we use two models of *Burton et al.* [1975] and *O'Brien and McPherron* [2000], which lead to almost similar results for this studied event. The observed *Dst* time profile is very well represented by the simulated *Dst*, giving a borderline moderate geomagnetic storm. The minimum *Dst* occurs with both models on July 14, 22 UT, with minima of −49 nT (Burton) and −57 nT (O'Brien). This is only 1 hour earlier compared to the observation, which peaks at −73 nT, and has thus a difference of only about 20 nT to the simulation. We have also added the Dst modeling using OMNI2 solar wind data (red and blue solid lines), which shows the Dst minimum to be underestimated at around -100 nT. Such a slight mismatch is often seen with the *Burton et al.* [1975] and *O'Brien and McPherron* [2000] models, which do not perfectly connect the L1 solar wind with the Dst observations.

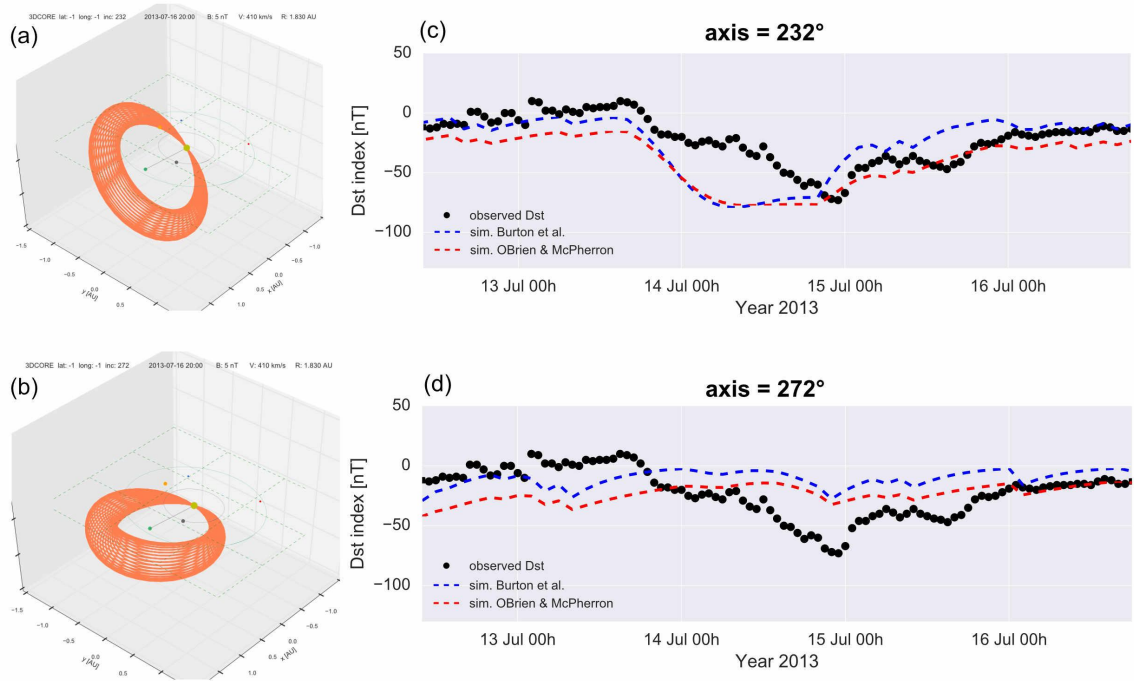

**Figure 7. Dst model sensitivity on flux rope axis orientation.** *Panels (a) and (b) show the torus visualization for two cases, where we have rotated the MFR axis orientation by - 20° (a) and + 20° away from the 252°* **as** *given by GCS modeling. Panels (c) and (d) show the resulting Dst profiles compared to the simulation, with the axis at 232° in (c) and at 272° in (d).* **The format is similar to Figure 6c.**

While these results seem promising, we clearly need to assess the sensitivity of the model output to the initial conditions. *Kay et al.* [2017] showed that their method of creating synthetic MFR observations is highly sensitive on the order of degrees for the orientation and direction of the CME. For our example, the possible range in the MFR direction longitude (latitude) leading to a hit at Earth is approximately -15° to +60° (-5° to +15°). Outside of this range, the MFR misses Earth. The difference for longitude and latitude is explained by the MFR inclination of 252° being only 18° away from a purely east-west oriented axis, thus a small change in latitude will lead to a miss quicker than a similar change in longitude because the flux rope is much wider in longitude in this case. However, both the timing and magnitude of the geomagnetic response changes widely over these possible ranges. We think that a complete sensitivity analysis for every simulation parameter is better suited for a CME MFR event which has been fully observed at two radially separated locations in the inner heliosphere [e.g. *Good et al.* 2015], but here we show the results of a first test concerning the dependence of *Dst* on flux rope orientation.

In **Figure 7** we show 3DCORE runs which differ only in the MFR axis orientation, keeping all other parameters similar. The results of the first run can be seen in Figure 7a and 7d. Just by tilting the axis by −20°, so the orientation is more inclined to the solar equator at 232° position angle, the *Dst* time profile strongly changes. This run is also closer to the orientation of the filament neutral line on the Sun (axis at 220°), i.e., what can be assumed to be the intrinsic orientation of the MFR. Whereas the *Dst* minimum is very close to the observed one, at around −75 nT, the simulated minimum (dashed lines) is now a plateau which reaches the *Dst* minimum more than half a day earlier compared to observation, and persists at peak values. In the second run in Figures 7b and 7d, the axis is tilted by +20° to a PA of 272°, so the flux rope axis is now almost perfectly aligned with the solar equatorial plane. Consequently, the resulting Dst is almost 0, clearly inconsistent with the observed values. These changes in *Dst* arise only from the change in synthetic *Bz* at Wind due to the different flux rope orientation, which also leads to slightly different trajectory through the synthetic flux rope. We leave more detailed analyses concerning the effects of variations in input parameters on the *Dst* results for future studies.

# 5 Conclusions

Forecasting the geomagnetic effects of solar storms (coronal mass ejections, CMEs) is currently massively hindered by our inability to predict the magnetic field configuration at the CME core, in particular their southward magnetic fields. Here, we have demonstrated how a new semi-empirical model (3DCORE) can be used to predict the speed and field components of a CME flux rope at 1 AU and derive the *Dst* index time series. The model is initiated by solar extreme ultraviolet and magnetogram observations, as well as coronagraph modeling results. Heliospheric Imaging on STEREO was used to confirm the unambiguous connection of the solar eruption to 1 AU.

In its first prototype, we additionally need in situ magnetic field data from along the Sun-Earth line, provided here by the MESSENGER spacecraft, to constrain 3 parameters of the MFR simulation − the magnitude of the axial magnetic field $B_0$, the flux rope diameter $D$ and $\Gamma$ parameter describing solar wind drag. Then, we propagate the model outward to 1 AU and find a good match between the synthetic and observed geomagnetic Dst index, based only the synthetic data input which includes a routine *HEEQ* to *GSM* coordinate conversion. Our simulation results conducted here can yield prediction lead times of 1 to 3 days, from the end of the flux rope observation at MESSENGER to its beginning and end at Wind. Our study may also be taken as a hint of the utility of having a solar wind monitor at or near the Sun-Earth line closer to the Sun than L1, but more studies are needed that use semi-empirical models such as presented here for *Dst* prediction, either only with solar and coronal inputs or with combined solar, heliospheric imaging and in situ inputs, to decide on the viability of such sub-L1 monitors for real time CME prediction in combination with physical modeling. We would also like to emphasize that even with very limited and sparse in situ information < 1 AU, a valuable constraint of CME propagation models should be possible. This is also supported by several studies showing that the deflection and rotation of CMEs largely takes place during about first 10% of their journey from the Sun to the Earth [e.g., *Isavnin et al.* 2014, *Kay et al.* 2016]. These observations also support the usage of 3DCORE for space weather forecasting. 3DCORE uses CME direction and orientation (with the axial field given by the source region) from the GCS reconstruction results around 20 solar radii away from the Sun, i.e., at the point where most dramatic change in CME geometrical parameters has already occurred.

The *Dst* results are sensitive to CME direction and orientation, thus very accurate results on these parameters are needed from observations [also shown by *Kay et al.* 2017]. This may be provided by a mission to the L5 point with multi-point coronagraph support (SOHO, STEREO) and a heliospheric imager capable of polarization measurements [*DeForest et al.* 2016]. We have also shown that the polarity inversion orientation of the source region is not a good predictor of the in situ orientation, and a rotation of about 30° is needed, which seems to have happened between the photosphere and the corona. Thus, we find further evidence that deflection and rotation of CMEs needs to be taken into account close to the Sun [e.g. *Kay et al.* 2016]. The point is that even though 30° inclination difference does not sound significant, this difference has a strong influence on the resulting magnetic field at L1 and consequently the predicted *Dst* index.

The processes of CME-CME interaction and merging are not included in the current 3DCORE version, and there is a reaction of the model only to a single-speed background solar wind, but not a 3D wind. The current 3 open parameters in the model should be based on CME statistics and other more sophisticated physical modeling regarding the origin, expansion and propagation of CMEs. The initial flux and helicity content of the flux rope may be set by examining the pre-eruptive state of the flux rope with coronal magnetic field modeling [e.g. *Yeates*, 2014; *Lowder and Yeates*, 2017] The circular cross-section and global shape also limits the current model applicability, which will be changed to different shapes such as ellipses or other deformed shapes [e.g. *Hidalgo et al.* 2002, *Owens et al.* 2006, *Janvier et al.* 2014, *Möstl et al.* 2015] in future updates.

We introduced in this short report the 3DCORE technique because it acts as part of a principle that we want to pursue in the following years in order to make advances on solving the *Bz* problem. The ultimate goal is to use solar, coronagraph and heliospheric imager observations in combination with L1 data to constrain the 3DCORE model parameter space in real time. A combination with the HI prediction model ElEvoHI [*Rollett et al.* 2016] could be a step in this direction. This model is based on HI observations and provides, besides the predictions of arrival time and speed, "side products" as the background solar wind speed, $\Gamma$ and kinematic profiles of the CME front. With 3DCORE, accurate L1 forecasts for up to 2 days in advance for the CME flux rope as the CME is sweeping over Earth could be possible. Essentially, even in real time, the shock arrival already constrains much of the CME kinematics, which has strong effects on the possible speeds and magnetic fields that arrive in the flux rope, given that the association with the remote observations is correct. 3DCORE is a technique that can easily produce many simulation runs quickly, and the runs that are expected to most accurately predict the current event can be selected by machine learning algorithms. These are based on the given solar and interplanetary inputs from many previous events and laws on the behaviour of CME magnetic flux ropes that are far from stochastic [e.g. *Bothmer and Schwenn* 1998, *Marubashi et al.* 2015, *Palmerio et al.* 2017]. The catalogs we have produced in *HELCATS* and others like the *Heliophysics Event Knowledgebase* are datasets that will be used to train these kind of algorithms. Essentially, for such a system it does not matter from where the in situ constraints come from, be it L1, which makes prediction lead time shorter, or in the far future from closer to the Sun by simple solar wind monitors near the Sun-Earth line, which, however, is expected to increase the prediction lead times considerably compared to L1.

Additionally, it has not escaped our notice that 3DCORE forms an approach that can produce synthetic in situ flux rope observations even when the spacecraft crosses the CME flux rope at a speed that is similar to the CME propagation speed, which could lead to interesting observational effects as the CME flux rope is not sampled along a 1D but essentially along a 3D trajectory. This will happen in the upcoming years likely a few times with observations of the *Parker Solar Probe* spacecraft. It will approach the Sun to about 0.05 AU, and spends enough time to likely observe a few CMEs at < 0.3 AU during the primary mission. Combined imaging and in situ observations by *Parker Solar Probe, Solar Orbiter* and *BepiColombo* will lead to further CME modeling constraints. 3DCORE and other semi-empirical models [*Isavnin*, 2016, *Kay et al.* 2017] will highly likely provide a valuable modeling context to interpret data returned by these new, groundbreaking missions.

## 6 Sources of data, codes and supplementary material

Catalogs used, described in Möstl et al. [2017]:

HIGeoCat: https://www.helcats-fp7.eu/catalogues/wp3_cat.html

Page for 2013 July 9:

CME: https://www.helcats-fp7.eu/catalogues/event_page.html?id=HCME_B__20130709_01

ARRCAT:

doi: 10.6084/m9.figshare.4588324

https://doi.org/10.6084/m9.figshare.4588324.v1

https://www.helcats-fp7.eu/catalogues/wp4_arrcat.html

ICMECAT:

doi: 10.6084/m9.figshare.4588315

https://doi.org/10.6084/m9.figshare.4588315.v1

https://www.helcats-fp7.eu/catalogues/wp4_icmecat.html

In situ data:

Wind: https://cdaweb.sci.gsfc.nasa.gov

MESSENGER: https://pds-ppi.igpp.ucla.edu

OMNI2 data for the *Dst* index:

description: https://omniweb.gsfc.nasa.gov/html/ow_data.html

data: ftp://nssdcftp.gsfc.nasa.gov/pub/data/omni/low_res_omni/omni2_all_years.dat

3DCORE code:

Möstl, C., 2017. doi: 10.6084/m9.figshare.5450341

https://doi.org/10.6084/m9.figshare.5450341

Animation of Figure 4:

https://doi.org/10.6084/m9.figshare.4602253

https://www.youtube.com/watch?v=Jr4XRzGCaaQ

Animation of Figure 5a:

https://figshare.com/s/b21f14d2098022689ada (this movie is part of the 3DCORE code)

## Acknowledgments, Samples, and Data

This study was supported by the Austrian Science Fund (FWF): [P26174-N27]. The presented work has received funding from the European Union Seventh Framework Programme (FP7/ 2007-2013) under grant agreement No. 606692 [HELCATS]. It was partially supported by NASA grant NNX16AO04G. This research has made use of SunPy, an open-source and free community-developed solar data analysis package written in Python [*The Sunpy Community et al.*, 2015]. It was carried out with the free ipython Anaconda environment using the packages numpy, scipy, matplotlib, pickle, sunpy and seaborn. We thank Alexis Rouillard, Brian Wood and Wenyuan Yu for discussions about coding 3D CME models.

All sources of data that were used in producing the results presented in this study are quoted in Section 6.

*Space Weather*

Supporting Information for

**Forward modeling of coronal mass ejection flux ropes in the inner heliosphere with 3DCORE**

**C. Möstl[1], T. Amerstorfer[1], E. Palmerio[2], A. Isavnin[2], C. J. Farrugia[3], C. Lowder[4], R.M. Winslow[3], J. M. Donnerer[1], E.K.J. Kilpua[2], P.D. Boakes[1]**

[1]Space ·Space Research Institute, Austrian Academy of Sciences, Graz, Austria.
[2]Department of Physics, University of Helsinki, Helsinki, Finland.
[3]Institute for the Study of Earth, Oceans, and Space, University of New Hampshire, Durham, NH, USA.
[4]Department of Mathematical Sciences, Durham University, Durham, United Kingdom.

## Contents of this file

## Introduction

This is the mathematical description of the 3DCORE model, including a table that clarifies the relationships between flux rope handedness, the axial field direction and expected magnetic fields components in various coordinate systems.

## Text S1.

### The *3-Dimensional COronal Rope Ejection* or 3DCORE model

In order to model the situation that is taking place during the observation of coronal mass ejection flux ropes in interplanetary space, we have created the 3DCORE model, which is an acronym for *3-Dimensional Coronal Rope Ejection*.

**Steps in 3DCORE**

1. **The flux rope shape:** As the general shape of the ICME MFR shell, we use an elliptical torus that becomes narrower at the ends where the MFR is bending towards the Sun [similar to *Hidalgo et al.* 2012]. Note that for the magnetic field configuration in this paper we use a circular cross-section, thus we put forward here the general ellipse equations only for a simpler future improvement, so we set the aspect ratio $a_r$ of the torus cross section to 1. We call this shape a *tapered torus*, which is described by the equations

(1)
$$
\begin{aligned}
a &= \rho_1(t), \\
b &= \rho_1(t)/a_r, \\
c &= 1, \\
f_{aux} &= c\sin(\psi/2), \\
x &= -(\rho_0(t) + a f_{aux}\cos(\phi))\cos(\psi), \\
y &= \rho_0(t) + a f_{aux}\cos(\phi)\sin(\psi), \\
z &= b f_{aux}\sin(\phi).
\end{aligned}
$$

Here, $\rho_0(t)$ is the major and $\rho_1(t)$ the minor radius of the torus, $f_{aux}$ the auxiliary function which describes how the torus narrows towards the Sun [*Hidalgo et al.* 2012], $\Psi$ are the coordinates along the torus axis and $\Phi$ the coordinate along its azimuthal direction, and *x,y,z* are the location of grid points of the CME in a cartesian coordinate system with the Sun in the origin, *x* positive towards the Earth, *y* towards solar west, and *z* normal to the solar equatorial *x-y* plane. These are *Heliocentric Earth Equatorial* (HEEQ) coordinates.

The torus center is at $\rho_0(t)$. To describe the propagation of the torus in heliocentric distance, we use the center of the cross section at the torus apex, which is situated at $2\rho_0(t)$. The torus axis thus forms a circle that always stays connected to the Sun.

The tapered torus shape can propagate in any direction, given by its HEEQ latitude *T* and longitude *P*. The torus is oriented at an angle $\eta$, the axis inclination angle, to the solar equatorial plane. The angle $\eta$ ranges from 0° to 360°, and increases in a counterclockwise direction. The value of the inclination $\eta = 0°$ means that the axis is pointing to solar north, at 90° it is pointing eastward, at 180° towards solar south and at 270° toward west, with north and south directions corresponding to normals to the solar equatorial plane (HEEQ). The flux rope axis direction is thus unambiguously defined, ranging from 0 to 360 degrees. Rotating the shape according to three angles is done with the Euler-Rodrigues relations. First, we rotate the torus by $P$ degrees round the $z$ axis, followed by a rotation of $T$ around a modified $y$-axis, and finally by a rotation of the angle $\eta$ around a modified $x$-axis. For the $T$ and $\eta$ rotations the $y$ and $x$ axes need to be changed according to spherical coordinates transformations, because after each applied rotation, the axis to which the rotation is applied changes.

2. **Adding the magnetic field:** We create a 3D MFR by placing 2.5D magnetic field cross sections along the MFR axis. Each cross section contains a grid of points, at steps in $d\Psi = 10°$, $d\Phi = 10°$ for the torus angles and $dr = 0.01$ AU in the radial direction in the torus cross section. In this first version of 3DCORE, we use circular-cross sections. However, due to the definition of the elliptical torus the model can be improved to include elliptical cross-sections in the future. Here, each of the circular torus cross-sections contains a 2.5D, uniform-twist Gold-Hoyle magnetic field [e.g. *Farrugia* et al. 1999, *Hu et al.* 2014], with $B_\psi$ the axial component and $B_\Phi$ the azimuthal component, and a non-existent radial field component:

$$B_\phi = \frac{H B_0(t) \tau r}{(1 + \tau^2 r^2)},$$
$$B_\psi = \frac{B_0(t)}{(1 + \tau^2 r^2)}, \qquad (2)$$
$$B_r = 0.$$

The twist $\tau$ is the number of turns a field line makes per AU around the axus, $r$ is the distance from the circle center, $B_0$ the axial field strength at the cross-section center, and $H$ is sign of the handedness or chirality. The twist parameter is in the range of 1 to 20 [see *Hu et al.* 2014]. The handedness $H$ and the inclination $\eta$ make it possible to describe all 8 ICME MFR types [*Bothmer & Schwenn* 1998, *Mulligan et al.* 1998], which includes right- and left-handed ropes at various inclinations. Table 1 shows the combinations of $\eta$ and $H$ to produce a desired MFR type and the expected magnetic field components for a direct (apex) impact.

3. **Propagating the flux rope:** The propagation of the torus away from the Sun is described with the drag-based-model [DBM, *Vrsnak* et al. 2013] by

$$r_{apex}(t) = \pm \frac{1}{\gamma} \ln[1 \pm \gamma(V_0 - w)t] + wt + R_0,$$

$$V_{apex}(t) = \frac{V_0 - w}{1 \pm \gamma(V_0 - w)t} + w. \qquad (3)$$

First, we define the apex position $r_{apex}(t)$ of the torus, which is the point with the furthest heliocentric distance. The movement of this point away from the Sun is described with a constant drag parameter $\gamma$ (in the paper text we use $\Gamma = \gamma \times 10^7$ km$^{-1}$) and w as the background solar wind speed. $V_{apex}(t)$ is the apex speed as function of time, and $V_0$ the CME initial speed at distance $R_0$, determined from coronagraph observations.

The torus expands quasi-linearly [e.g. *Leitner et al.* 2007], with $D_{1AU}$ being the MFR diameter at 1 AU, which is constrained by MESSENGER observations in such a way that the simulation magnetic field time series is consistent with the diameter at MESSENGER.

$$\rho_1(t) = \frac{D_{1AU}[r_{apex}(t)]^{1.14}}{2},$$

$$\rho_0(t) = \frac{r_{apex}(t) - \rho_1(t)}{2} \qquad (4)$$

Equations (4) means that we first determine the torus minor radius $\rho 1$, and then the position of the torus center at $\rho 0$, which follows from the definition of the apex distance $r_{apex}(t) = 2\rho 0(t) + \rho 1(t)$.

Th axial magnetic field decreases as a power law, with the exponent of -1.64, also taken from *Leitner et al.* [2007],

$$B_\rho(t) = B_{1AU}[2\rho_0(t)]^{-1.64}. \qquad (5)$$

This equation (5) relates the magnetic field at 1 AU ($B_{1AU}$) to the axial magnetic field $B_r$ at the current heliocentric distance of the torus axis, which is positioned at $2\rho_0(t)$.

4. **Making synthetic measurements:** A spacecraft can be placed at any position in the model heliosphere to extract synthetic in situ data along a given spacecraft position, given by latitude, longitude and heliocentric distance in HEEQ. As the CME expands and propagates into the heliosphere, at each timestep it is checked where the next grid point of the CME MFR is situated with respect to the spacecraft position. This needs to be defined because the spacecraft is of course never exactly on a position of a grid point in the CME MFR. To this end, the measurement parameter $m$ is defined: for a synthetic measurement to be taken, $m < 0.05$ AU. Note that the choice of $m$ is related to the grid resolution.

The last step of 3DCORE consists in the conversion of the MFR magnetic field components into the coordinate system of the spacecraft, which are often given in the RTN system (STEREO, MESSENGER), and for Earth the GSE and GSM systems are needed for comparison to magnetospheric indices. In the paper, however, we have used data converted to HEEQ and and SCEQ for easier comparison.

The unit vectors of the points in the torus are in cartesian coordinates given by:

$$\vec{e}_r = [\cos(\phi)\cos(\psi), \cos(\phi)\sin(\psi), \sin(\phi)]$$
$$\vec{e}_\phi = [-\sin(\phi)\cos(\psi), -\sin(\phi)\sin(\psi), \cos(\phi)] \qquad (6)$$
$$\vec{e}_\psi = [-\sin(\psi), \cos(\psi), 0]$$

The magnetic field components in the torus coordinates and torus cartesian coordinates are given as (see equation 2):

$$\vec{B} = [B_r \vec{e}_r, B_\phi \vec{e}_\phi, B_\psi \vec{e}_\psi] = [B_x \vec{e}_x, B_y \vec{e}_y, B_z \vec{e}_z]. \qquad (7)$$

The components $Bx$, $By$ and $Bz$ can be found by the dot product with the cartesian unit vectors

$$B_x = B_r \vec{e}_r \cdot \vec{e}_x + B_\phi \vec{e}_\phi \cdot \vec{e}_x + B_\psi \vec{e}_\psi \cdot \vec{e}_x \qquad (8)$$
$$B_y = B_r \vec{e}_r \cdot \vec{e}_y + B_\phi \vec{e}_\phi \cdot \vec{e}_y + B_\psi \vec{e}_\psi \cdot \vec{e}_y$$
$$B_z = B_r \vec{e}_r \cdot \vec{e}_z + B_\phi \vec{e}_\phi \cdot \vec{e}_z + B_\psi \vec{e}_\psi \cdot \vec{e}_z$$

The results of these calculations are

$$B_x = B_r \cos(\phi)\cos(\psi) - B_\phi \sin(\phi)\cos(\psi) - B_\psi \sin(\psi),$$
$$B_y = B_r \cos(\phi)\sin(\psi) - B_\phi \sin(\phi)\sin(\psi) + B_\psi \cos(\psi),$$ (9)
$$B_z = B_r \sin(\phi) + B_\phi \cos(\phi).$$

Hence, we have the magnetic field components available in a cartesian system, but we have not yet rotated the cloud accordingly to its propagation direction and inclination. Thus, we now rotate this magnetic field vector ($B_x$, $B_y$, $B_z$) according to the Euler-Rodrigues relations, similar to point 2 above for the CME shape. Finally, these components need to be projected into coordinate system at a spacecraft to be comparable to real observations. For calculating the Dst index at Earth, the GSM system is used, which is done following the procedures in *Hapgood* [1992]. From HEEQ, a conversion to HAE and HEE is done. GSE is then produced by a sign reversal in HEE and the final GSM components are calculated from the GSE values.

The bulk plasma speed of the point that is measured inside the CME can be calculated from $V_{apex}(t)$ by a simple argument involving self-similarity: because the geometry of the CME shape does not change with time (only its size changes), the relations [*Möstl and Davies*, 2013] are valid:

$$\frac{r_{apex}(t)}{V_{apex}(t)} = \frac{r_p(t)}{V_p(t)},$$ (10)
$$V_p(t) = r_p(t) V_{apex}(t) / r_{apex}(t).$$

Here, $r_p(t)$ and $V_p(t)$ are the heliocentric distance of the point where the magnetic field is measured.

**Table S1.** Relation of CME flux rope type to inclination angle $\eta$ and handedness $H$ in 3DCORE. Type: N = north, W = west, S = south, E = east. $H$: Left-handed = -1, right-handed = +1.

|            | SEN    | SWN    | NES    | NWS    | WNE    | ESW    | ENW    | WSE    |
|------------|--------|--------|--------|--------|--------|--------|--------|--------|
| H          | -1     | 1      | 1      | -1     | 1      | 1      | -1     | -1     |
| $\eta$     | 90     | 270    | 90     | 270    | 0      | 180    | 0      | 180    |
| $B$ (RTN, SCEQ) | $-B_N$ | $-B_N$ | $+B_N$ | $+B_N$ | $+B_T$ | $-B_T$ | $-B_T$ | $+B_T$ |
|            | $-B_T$ | $+B_T$ | $-B_T$ | $+B_T$ | $+B_N$ | $-B_N$ | $+B_N$ | $-B_N$ |
|            | $+B_N$ | $+B_N$ | $-B_N$ | $-B_N$ | $-B_T$ | $+B_T$ | $+B_T$ | $-B_T$ |
| $B$ (GSE, GSM) | $-B_z$ | $-B_z$ | $+B_z$ | $+B_z$ | $-B_y$ | $+B_y$ | $+B_y$ | $-B_y$ |
|            | $+B_y$ | $-B_y$ | $+B_y$ | $-B_y$ | $+B_z$ | $-B_z$ | $+B_z$ | $-B_z$ |
|            | $+B_z$ | $+B_z$ | $-B_z$ | $-B_z$ | $+B_y$ | $-B_y$ | $-B_y$ | $+B_y$ |